# Unraveling the nature of thermally induced spin reorientation in $NdFe_{1-x}Cr_xO_3$

Jiyu Shen[1,4], Jiajun Mo[1,4], Zeyi Lu[1], Huilin Huang[1], Chenying Gong[1], Yi Long [1], Arramel[3], Yanfang Xia*,[1], Min Liu*,[1], Jizhou Jiang*,[2]

[1]*College of Nuclear Science and Technology, University of South China, Hengyang 421200, Hunan, P.R China.*
[2]*School of Environmental Ecology and Biological Engineering, School of Chemistry and Environmental Engineering, Key Laboratory of Green Chemical Engineering Process of Ministry of Education, Wuhan Institute of Technology, Wuhan 430205, P.R China.*
[3]*Nano Center Indonesia, Jalan Raya PUSPIPTEK, South Tangerang, Banten 15314, Indonesia.*
[4]*These authors contributed equally to this work.*
*Email: xiayfusc@126.com, liuhart@126.com, 027wit@163.com*

**Abstract**

Understanding spin control mechanisms is an important part of condensed matter physics and the theoretical basis for designing spintronic devices. In this letter, based on four-sublattices molecular field theory, we propose that the underlying $NdFe_{1-x}Cr_xO_3$ magnetic mechanism is driven by spin reorientation sensitive to temperature. The actual coupling angular momentum, angle between the $Nd^{3+}$ and $Cr^{3+}/Fe^{3+}$ moments at the given temperature is realized via the $Nd^{3+}$ magnetic moment projection onto the $Cr^{3+}/Fe^{3+}$ plane. As the temperature increases, the angle between the moment of $Nd^{3+}$ and the moment of $Cr^{3+}/Fe^{3+}$ decreases monotonically. In this work, the magnetic mechanism of $NdFe_{1-x}Cr_xO_3$ ($x$=0.1, 0.9), the close relationship between A/B angle and temperature are presented, which laid a theoretical foundation for the design of new multifunctional magnetic materials.

# 1. Introduction

Ultra-fast spin control is not only the main direction of future spintronic and information storage technology development, but also the basis for designing nanoscale spintronic devices [1]. It has been widely used in spin torque device components [2]. In previous reports, various methods have been developed to achieve this control. For instance, precise control methods for femtosecond lasers with high output power [3-5] and heat-induced control [6,7]. Rare earth perovskites are one of the materials with typical thermally induced control possibilities. In comparison to ordinary perovskite, the coupling of magnetic rare-earth ions at the A and B sites significantly improve the overall magnetism of the material and enables a variety of applications, including magneto-optical materials [8], catalysts [9,10], and sensors [11,12].

In rare-earth perovskites, spin reorientation occurs between two long-range magnetic orders and originates from 3d-4f interactions. It can lock the 3d magnetic moment orientation in the anisotropic direction of the interface formed by the magnetic moment direction of rare-earth ions, and is temperature-dependent. In order to achieve better spin control, many scholars have made outstanding contributions to the spin reorientation of different rare-earth perovskites. Vedmedenko et al. [13] theoretically investigate spin-reorientation scenarios in 2D thin films within a first-order anisotropy approximation by means of Monte Carlo simulations. Another paradigm for spin reorientation in twisted perovskites ($R_2CuMnMn_4O_{12}$ ($R$=Y or Dy)) is shown by Vibhakar et al. [14]: Triple A-site columnar ordered quadruple perovskites with three ordered magnetic phases and up to two spin-redirected phase transitions.

Furthermore, molecular field theory is a first-order approximation of the Heisenberg Hamiltonian, which simplifies the space relationship of interaction while accurately describing the general law of magnetic phase transitions. The majority of reports have used molecular field theory to predict the magnetic origin of various materials and obtained very good verification results [15,16], indicating that molecular field theory is an extremely effective method for dealing with complex systems.

In this letter, we referred to the method of previous researchers and synthesized $NdFe_{1-x}Cr_xO_3$ samples by sol-gel combustion method [17] and performed a series of characterization tests, such as X-ray diffraction, Fourier transform infrared spectroscopy, Mössbauer spectroscopy, magnetic measurements. The results of structural characterization showed that single-phase pure $NdFe_{1-x}Cr_xO_3$ was synthesized. The macroscopic magnetism of $NdFe_{1-x}Cr_xO_3$ was theoretically investigated by means of magnetic analysis and molecular field theory. A precise exchange between the A and B sites mediated the influence of spin reorientation in this system is resolved.

## 2. Sample Preparation

$NdFe_{1-x}Cr_xO_3$ ($x$=0.1, 0.9) is prepared by a simple sol-gel combustion method. The required precursors are all from Macklin company, namely neodymium oxide ($Nd_2O_3$), ferric nitrate nonahydrate ($Fe(NO_3)_3 \cdot 9H_2O$), chromium nitrate nonahydrate ($Cr(NO_3)_3·9H_2O$), nitric acid ($HNO_3$), ethylene glycol ($C_2H_6O_2$), citric acid ($C_6H_8O_7$). First, the neodymium oxide was dissolved with excess nitric acid to obtain a neodymium nitrate solution. Then, other nitric acid compounds in water were mixed with neodymium nitrate solution according to a certain ionic ratio ($Nd^{3+}$: $Fe^{3+}$: $Cr^{3+}$=1: 1-$x$: $x$, $x$=0.1, 0.9). An excess of citric acid and a certain amount of ethylene glycol were combined as the mixed solution molar ratio of 1:1. The solution was stirred evenly at 80°C on a magnetic stirrer until a gel formation. Then, the gel was heated at 120 °C. The obtained powder was pre-calcined at 600 °C for 12 h and subsequently calcined at 1200 °C for 24 h, upon cooling, the final nanopowder samples were obtained.

## 3. Characterizations

We not only performed magnetic tests on the samples using vibrating sample magnetometers (VSM), but also carried out structural characterizations, such as: X-ray diffraction (XRD), Fourier infrared spectroscopy (FT-IR), Mössbauer spectroscopy and magnetic test. The test results can be seen in Fig. 1-5.

## 4. Results and discussion

### 4.1. XRD Analysis

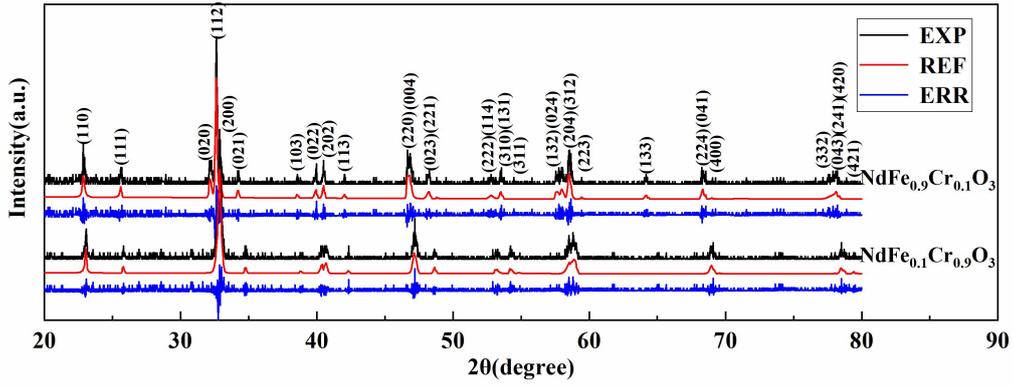

**Fig. 1** X-ray diffraction pattern of NdFe$_{1-x}$Cr$_x$O$_3$ ($x$=0.1 and 0.9).

The XRD results of NdFe$_{1-x}$Cr$_x$O$_3$ ($x$=0.1 and 0.9) samples were analyzed using FullProf software, and the results are presented in Fig. 1. The black line represents the experimental test data, the blue line represents the data after software refinement, and the red line is the error between the two. The prominent diffraction peaks of the samples are marked on their respective signals. It can be judged from the main Bragg peaks that all samples have orthogonal space groups. Besides, the XRD pattern of the NdFe$_{0.9}$Cr$_{0.1}$O$_3$ sample has a very obvious peak splitting phenomenon at 32°~34° in Fig. 1. This shows that the perovskite structure undergoes orthogonal distortion under the stress contribution. The structural parameter ($D_{OD}$) is introduced to characterize the orthogonal distortion in the following formula: [18] For the lattice of the *Pbnm* space group:

$$D_{OD} = \frac{1}{3}\sum_{i=1}^{3}|\frac{a_i - \bar{a}}{\bar{a}}| * 100\% \quad (1)$$

For the lattice of the *Pbnm* space group, $\alpha_1 = a$, $\alpha_2 = b$, $\alpha_3 = c/\sqrt{2}$, $\bar{\alpha} = (a * b * c/\sqrt{2})^{\frac{1}{3}}$.

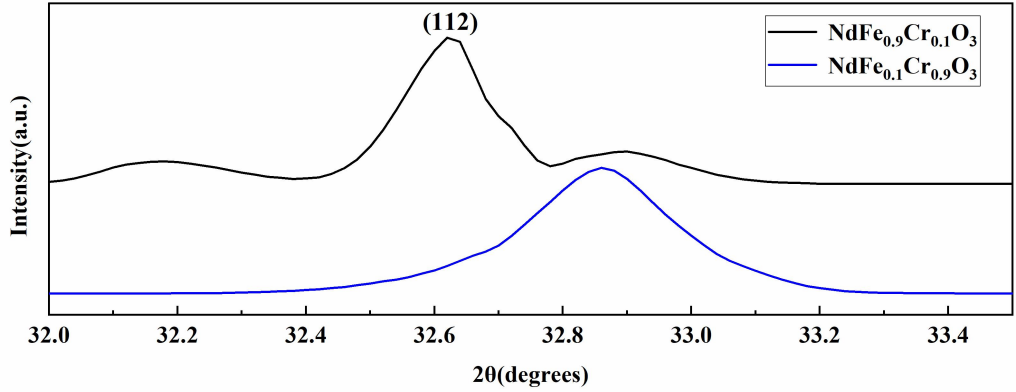

**Fig. 2** The enlarged image of NdFe$_{1-x}$Cr$_x$O$_3$ ($x$ = 0.1 and 0.9) sample near the diffraction peak (112).

Fig. 2 shows the enlarged region in which a broadened diffraction peak is plotted as the function of Fe content. It turns out that central peak belongs to (112) reflection plane is shifted to lower angles. This phenomenon is attributed to the increased shrinkage of the stress gradient that commonly found in polycrystalline semiconductor compounds [19]. The crystal stress is expressed by the Williamson-Hall (W-H) method, i.e., formula (2) [20, 21]:

$$\beta \cos\theta = k\lambda/D + 4\varepsilon \cos\theta \quad (2)$$

Where $k$ = 0.9, $\lambda$ is the X-ray wavelength (1.5418 nm), $\beta$ is the full width at half maximum of the peak, $D$ is the average crystallite size, $\theta$ is the Bragg angle, and $\varepsilon$ is the macrostrain.

**Table. 1** Lattice parameters, unit cell volume, average grain size and bulk density data for $NdFe_{1-x}Cr_xO_3$ ($x = 0.1$ and 0.9).

| Sample | $a$ (Å³) | $b$ (Å³) | $c$ (Å³) | Cell volume (Å³) | Average grain size (Å) | Bulk density (g/cm³) |
|---|---|---|---|---|---|---|
| $NdFe_{0.9}Cr_{0.1}O_3$ | 5.453 | 5.584 | 7.768 | 236.53 | 725 | 6.9664 |
| $NdFe_{0.1}Cr_{0.9}O_3$ | 5.422 | 5.478 | 6.692 | 228.54 | 529 | 7.0980 |

The lattice parameters, unit cell volume, average grain size, and other parameters of the $NdFe_{1-x}Cr_xO_3$ ($x = 0.1$ and 0.9) samples can be seen in Table. 1. Here we found that the lattice parameters, unit cell volume, and average grain size increased significantly as the function of iron content, caused by the difference in the ionic radius of $Fe^{3+}$ (0.645 Å) and $Cr^{3+}$ (0.615 Å) [22].

### 4.2. FT-IR Analysis

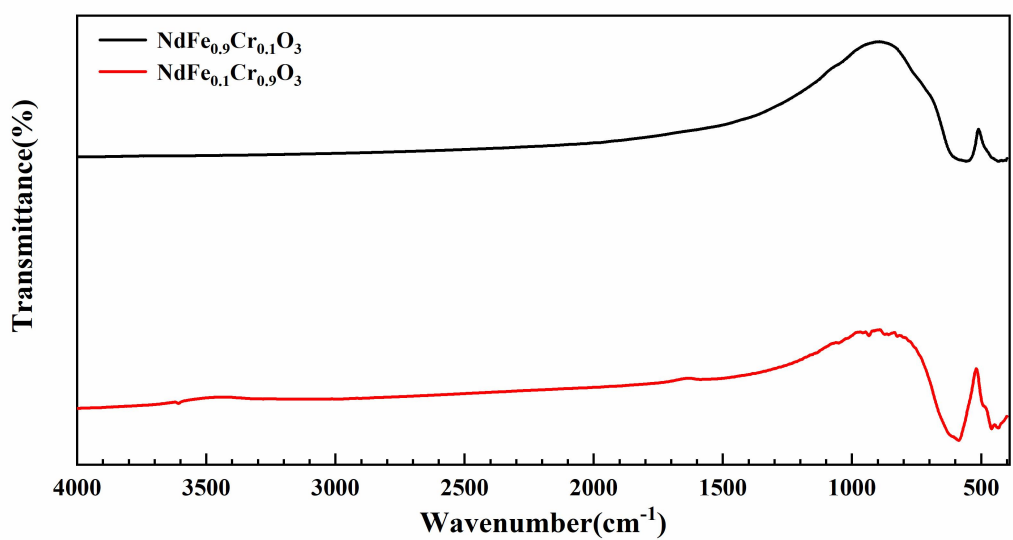

**Fig. 3** Fourier transform infrared spectroscopy of $NdFe_{1-x}Cr_xO_3$ ($x = 0.1$ and 0.9).

The FT-IR spectra of NdFe$_{1-x}$Cr$_x$O$_3$ ($x$ = 0.1 and 0.9) are shown in Fig. 3. Both samples show two distinct characteristic absorption peaks, located between 400 cm$^{-1}$-500 cm$^{-1}$ and 500 cm$^{-1}$-600 cm$^{-1}$, respectively. The absorption peaks in the lower wavenumber range are caused by the vibration of the O-Cr/Fe-O bond [23]. The absorption peaks in the higher wavenumber range correspond to the vibration of the Cr/Fe-O bond. As the Fe content increases, the absorption peak is shifted to the low wavenumber region, indicating that Fe ions have a good substitution effect on Cr$^{3+}$ ions.

### 4.3. Mössbauer Spectral Analysis

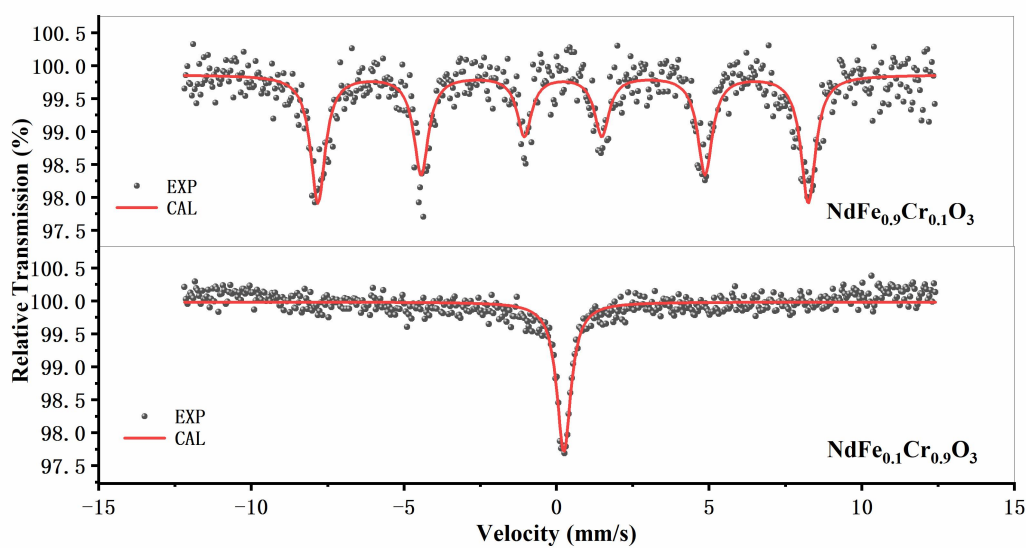

**Fig. 4** Mössbauer spectrum of NdFe$_{1-x}$Cr$_x$O$_3$ ($x$ = 0.1 and 0.9) at 300 K.

| Sample | T (K) | QS (mm/s) | IS (mm/s) | H (T) | Γ (mm/s) | Area (%) |
|---|---|---|---|---|---|---|
| NdFe$_{0.9}$Cr$_{0.1}$O$_3$ | 300K | 0.00 | 0.227 | 49.982 | 0.582 | 100 |
| NdFe$_{0.1}$Cr$_{0.9}$O$_3$ | 300K | \ | 0.220 | \ | 0.323 | 100 |

**Table. 2** Hyperfine parameters of NdFe$_{1-x}$Cr$_x$O$_3$ ($x$ = 0.1 and 0.9) samples.

The Mössbauer spectra of NdFe$_{1-x}$Cr$_x$O$_3$ ($x$ = 0.1 and 0.9) were measured at 300 K and parameters such as isomer shifts (IS), quadrupole splitting (QS), and hyperfine field (H) were fitted using MössWinn 4.0 software. Table. 2 contains all fitted data. Fig. 4 shows the Mössbauer spectra fit result. IS and QS values indicating the existence of Fe$^{3+}$ where no divalent or tetravalent iron were found. It is worth noting that the spectral line at $x$ = 0.1 is a standard Zeeman Sextet, indicating that the system poses an intense exchange coupling interaction. Since the measuring condition at room temperature is much lower than the phase transition temperature, we tentatively propose that this system contains single pure phase NdFe$_{0.9}$Cr$_{0.1}$O$_3$. The QS value is nearly zero, indicating that the electric field gradient around the Fe atoms is zero, Cr$^{3+}$ ions are dispersed uniformly. Thus, the polarization effect of Cr-O- is negligible. The intensity of this hyperfine field is much larger than that of BiFe$_{0.9}$Cr$_{0.1}$O$_3$ [24], indicating the introduction of Nd$^{3+}$ ions increased significantly the coupling interaction of B site ions, and the phase transition temperature is higher than 600 K (compared to BiFe$_{0.9}$Cr$_{0.1}$O$_3$). In the case of $x$ = 0.9, it contains a low proportions Fe$^{3+}$ ions and consequently this particular compound is difficult to form Fe-Fe clusters. Fe is usually coordinated by Fe-O-Cr, and all Cr$^{3+}$ ions in the vicinity of Fe$^{3+}$ ions should exhibit same polarization effect on the electron cloud of Fe$^{3+}$ ions, resulting in an electric field gradient that led to a singlet configuration.

### 4.4. Magnetization Curve Analysis

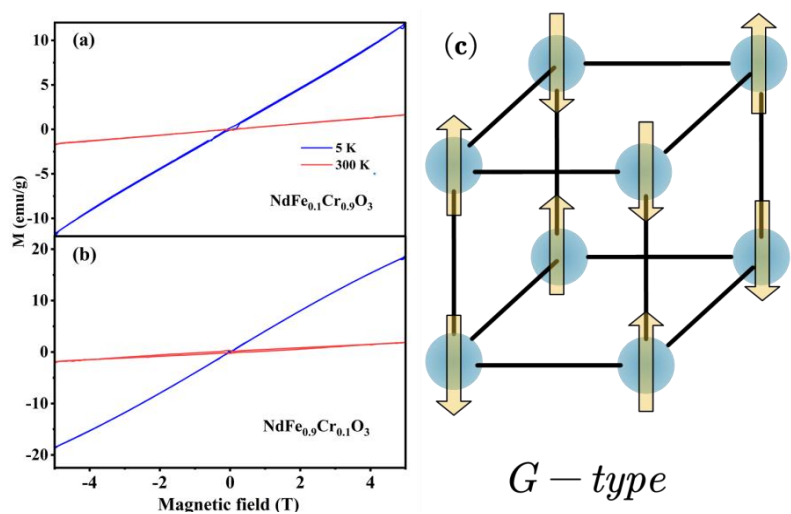

**Fig. 5** (a) The M vs H curves of NdFe$_{0.1}$Cr$_{0.9}$O$_3$ tested at 5 K and 300 K. (b) The M vs H curves of NdFe$_{0.9}$Cr$_{0.1}$O$_3$ tested at 5 K and 300 K. (c) The antiferromagnetic configuration of the system.

The magnetization curves (M-H) of the NdFe$_{1-x}$Cr$_x$O$_3$ system at 300 K and 5 K are shown in Fig. 5(a-b). Except for NdFe$_{0.9}$Cr$_{0.1}$O$_3$ at 300 K, the other compounds displayed straight lines with varying slopes. We estimate the remanence (0.33 emu/g) and coectivity (5000 Oe) of NdFe$_{0.9}$Cr$_{0.1}$O$_3$ at 300 K via Fig. 5a, which depicts the system with weak ferromagnetic behaviour. Under external magnetic fields of up to 6 T, no samples reach saturation magnetization, and the magnetization at low and high temperatures is quite different. The two samples maintain antiferromagnetic order at 5 K. This can be understood due to the superexchange interaction (Cr$^{3+}$-O-Cr$^{3+}$) that arises by the introduced Cr$^{3+}$ ions, i.e., there are antiparallel spin direction between adjacent ions in structure i.e., G-type antiferromagnetic configuration, as shown in Fig. 5c. The magnetization of $NdFe_{0.1}Cr_{0.9}O_3$ is significantly lower than that of $NdFe_{0.9}Cr_{0.1}O_3$, which is because $NdCrO_3$ has weaker antiferromagnetic properties than $NdFeO_3$.

**4.5. Thermomagnetic Curve Analysis and Simulation**

We analyzed the macroscopic magnetic characteristics of the NFCO system at temperatures of 300 K and 5 K (Fig. 5(a-b)). Lines with different slopes (except $NdFe_{0.9}Cr_{0.1}O_3$ at 300 K, which exhibit weak ferromagnetic behavior) show strong antiferromagnetic behavior of the system. The G-type antiferromagnetic configuration (Fig. 5c), i.e. the anti-parallel alignment of adjacent ion spins, is the main cause of the macroscopic magnetism of the system.

Figure 6a shows the magnetization *vs* temperature curve over the temperature range of 5 K to 400 K. The paramagnetic phase transition temperature of $NdFe_{0.1}Cr_{0.9}O_3$ is observed at about 225 K, but no phase transition of $NdFe_{0.9}Cr_{0.1}O_3$ is observed before 400 K. This is consistent with the conclusions obtained in the M-H curve. Below 225 K, the remaining electron and spin-orbit coupling induces superexchange interaction facilitating the spin angle of the $Cr^{3+}$ ion reduction to slightly less than 180°, and the weak ferromagnetic behavior is recorded below this temperature [25]. Multiple magnetic phase transitions are frequently observed in perovskites made of rare earth ortho-chromium or rare earth ferrite [26,27]. For the $NdFe_{1-x}Cr_xO_3$ system, the phase transition temperature observed at high temperatures, interestingly, a distinct magnetization transition temperature is also observed at a low temperature, which is commonly referred to as the spin reorientation temperature ($T_{SR}$). We attribute the finding can be associated to the competition between A-site rare-earth ions (Eu, Nd, La, and so on) and B-site metal ions. Here, $Nd^{3+}$ at A site is experiencing superexchange interaction with the $Cr^{3+}/Fe^{3+}$ cation at B site, thus strong effective fields polarize $Nd^{3+}$ cation and generate the spin direction of the $Cr^{3+}/Fe^{3+}$ ion [28,29]. The spin of $Cr^{3+}$ ion rotates from Γ2 (Fx, Cy, Gz) to Γ1 (Ax, Gy, Cz), while the spin of $Fe^{3+}$ is more complex, from Γ4 (Gx, Ay, Fz) to Γ1 (Ax, Gy, Cz), passing through Γ24 (Gxz, Fxz) and Γ2 (Fx, Cy, Gz). As the temperature decreases, this spin transition can be reflected in Fig. 7. Figure 6c illustrates the Curie-Weiss law ($\chi = \frac{C}{T-\theta}$) fit result of $\chi^{-1}$ for $NdFe_{0.1}Cr_{0.9}O_3$, where C is the Curie constant and θ is the Weiss constant. We determined that the Curie constant C is approximately 4.265 emu·K/Oe·mol and the Weiss constant is approximately -203.6 K. The negative Weiss value indicates that antiferromagnetic coupling dominates the system. The effect moment, $\mu_{eff} \sim 5.84$ $\mu_B$, is consistent with the theoretical value of 5.66 $\mu_B$ ($S_{Nd}$ = 3/2, $S_{Fe}$ = 5/2, $S_{Cr}$ = 3/2). The small difference is primarily due to the absence of orbital angular momentum, and consequently theoretical value's lande factor could not be 2.

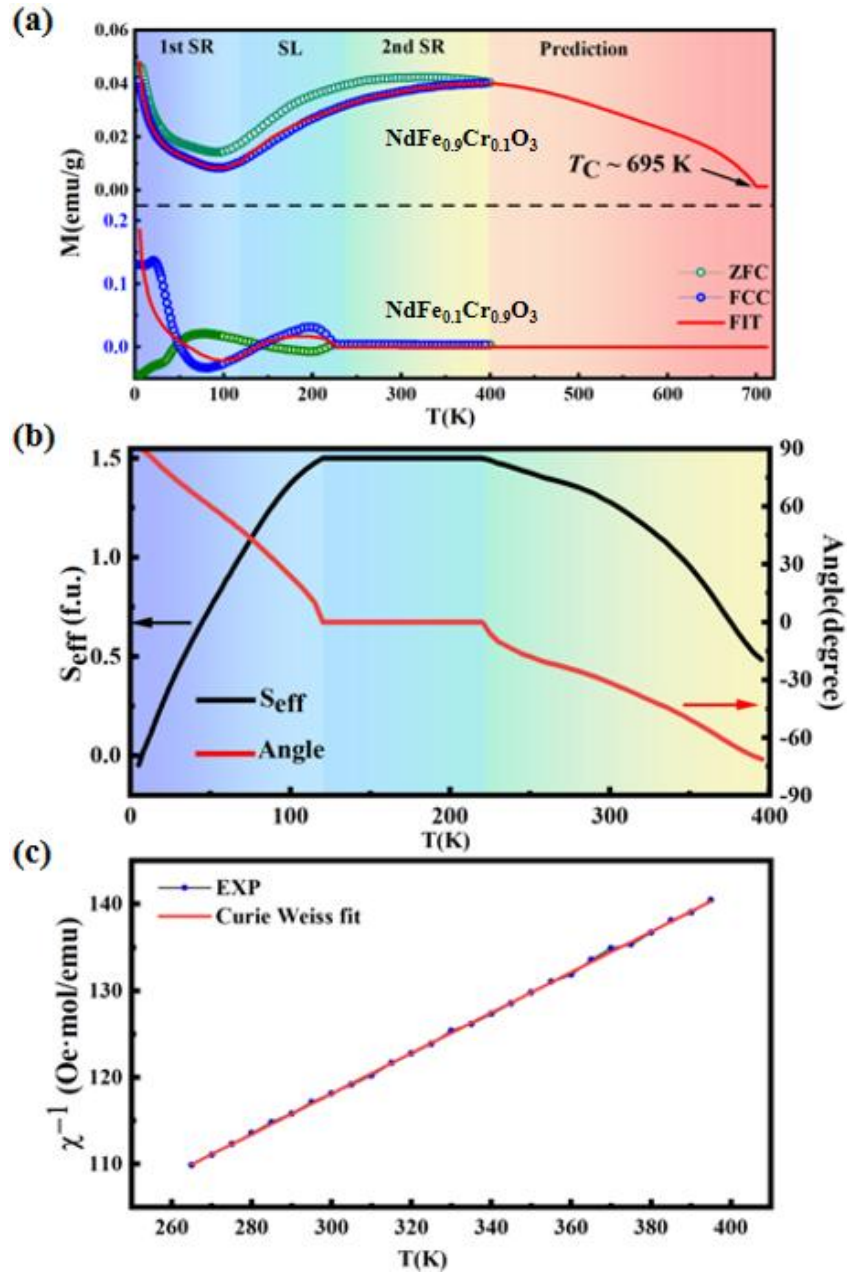

**Fig. 6** (a) The *M vs T* curves of NFCO under external magnetic fields of 100 Oe. The molecular field theory calculation and high temperature prediction curve of $NdFe_{0.9}Cr_{0.1}O_3$ and $NdFe_{0.1}Cr_{0.9}O_3$. (b) The effect spin of $Nd^{3+}$ ions and the angle variation curve between $Nd^{3+}$ and $Fe^{3+}/Cr^{3+}$ ions. (c) The fitting result of Curie Weiss law in $NdFe_{0.1}Cr_{0.9}O_3$.

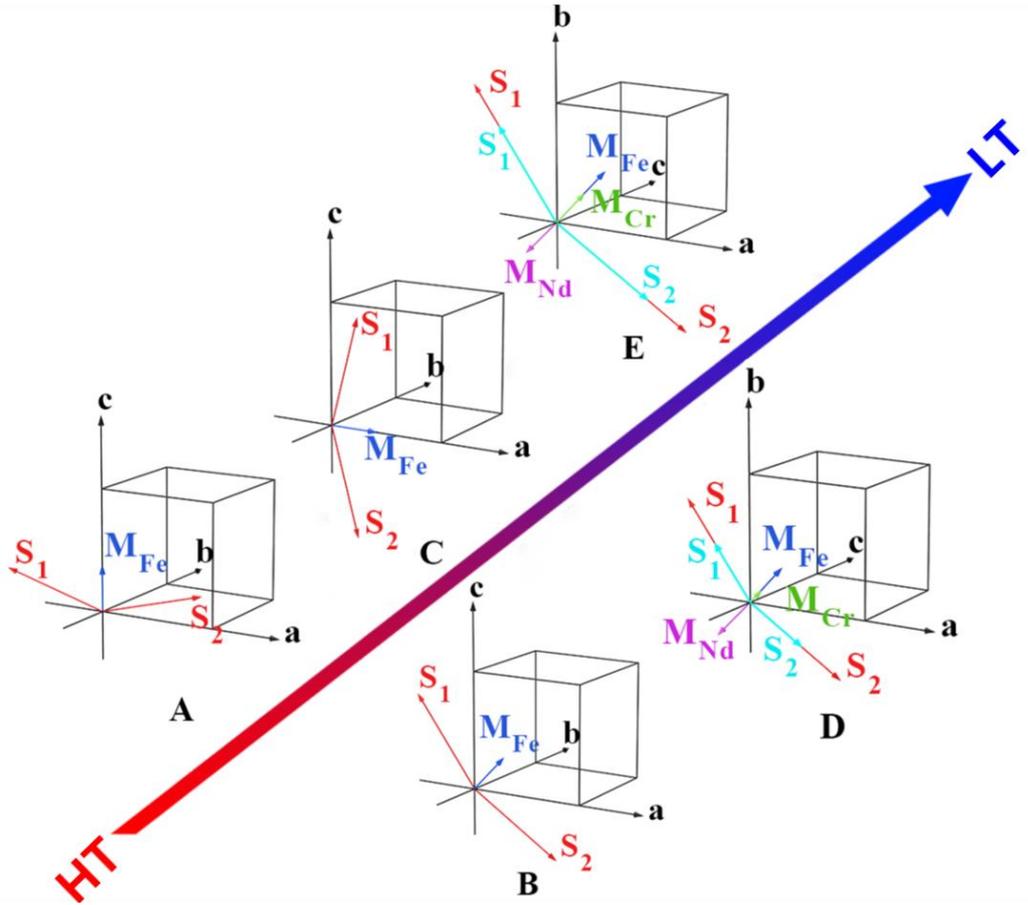

**Fig. 7** The magnetic moment contribution and spin reorientation of each ion in the NdFe$_{1-x}$Cr$_x$O$_3$ system as a function temperature. (HT: high temperature, LT: low temperature. The long black arrows at the bottom point in the direction of decreasing temperature; the red and aqua-green lines represent a pair of spins of Fe$^{3+}$/Cr$^{3+}$ ions, respectively; the blue, green and purple lines represent the magnetic contributions of Fe$^{3+}$/Cr$^{3+}$/Nd$^{3+}$, respectively. A, B, C, D, E diagrams represent the Γ4, Γ24, Γ2, Γ1, Γ1` spin structures, respectively).

In order to understand the magnetic mechanism of system, we consider a relatively straightforward method to uncover the magnetic properties, namely, the molecular field theory. The NdFe$_{0.1}$Cr$_{0.9}$O$_3$ system can be described by four sublattices, denoted by the letters L$_{Fe}$, L$_{Cr}^a$, L$_{Cr}^b$, and L$_{Nd}$, correspond to the sublattices decorated by the $Cr^{3+}$ ions in the a- and b-sites, respectively. We disregard the Nd-Nd superexchange in this work because it is too insignificant to have an effect on the system. Each of the sublattice's mean-field can be expressed as:

$$\begin{aligned}
H_{Fe} &= \lambda_{FeCr^a} M_{Cr^a} + \lambda_{FeNd} M_{Nd} + h \\
H_{Cr^b} &= \lambda_{Cr^b Cr^a} M_{Cr^a} + \lambda_{Cr^b Nd} M_{Nd} + h \\
H_{Cr^a} &= \lambda_{Cr^a Cr^b} M_{Cr^b} + \lambda_{Cr^a Fe} M_{Fe} + \lambda_{Cr^a Nd} M_{Nd} + h \\
H_{Nd} &= \lambda_{NdFe} M_{Fe} + \lambda_{NdCr^b} M_{Cr^b} + \lambda_{NdCr^a} M_{Cr^a} + h
\end{aligned} \quad (3)$$

For $NdFe_{0.9}Cr_{0.1}O_3$, the mean-field can be expressed as:

$$\begin{aligned}
H_{Cr} &= \lambda_{CrFe^a} M_{Fe^a} + \lambda_{CrNd} M_{Nd} + h \\
H_{Fe^b} &= \lambda_{Fe^b Fe^a} M_{Fe^a} + \lambda_{Fe^b Nd} M_{Nd} + h \\
H_{Fe^a} &= \lambda_{Fe^a Fe^b} M_{Fe^b} + \lambda_{Fe^a Cr} M_{Cr} + \lambda_{Fe^a Nd} M_{Nd} + h \\
H_{Nd} &= \lambda_{NdCr} M_{Cr} + \lambda_{NdFe^b} M_{Fe^b} + \lambda_{NdFe^a} M_{Fe^a} + h
\end{aligned} \quad (4)$$

Where $\lambda_{ij}$ represents the molecular field constant between $i$ and $j$ sublattices, and it is proportional to the exchange constant $J_{ij}$, $h$ is the external field; $M_i$ is the magnetization of $i$ sublattice. The magnetization of $i$ sublattice is:

$$M_i = x_i N_A g \mu_B S_i B_{Si}\left(\frac{g \mu_B S_i H_i}{k_B T}\right) \quad (5)$$

Where $x_i$ is the molar quantity of $i$ ions, $g$ is the lande factor, and $\mu_B$ represents the Bohr magneton. $N_A$ is the Avogadro constant. $S_i$ is the spin quantum number of $i$ ions ($S_{Fe} = 5/2$, $S_{Cr} = 3/2$, $S_{Nd} = 3/2$). The exchange constant $J_{MM}$ between $M$ and $M$ ions can be calculated by:

$$|J_{MM}| = \frac{2 Z_{MM} S_M (S_M + 1)}{3 k_B T_N^M} \quad (6)$$

Where $Z_{MM}$ is the number of $M$ ions required to be $M$ ions nearest neighbours. $k_B$ is the Boltzmann constant, and $T_N^M$ is the Phase transition temperature of $NdMO_3$. The exchange constants between Fe and Fe ($J_{FeFe}$) and Cr and Cr ($J_{CrCr}$) have been calculated to be 14.67 and 20.0 K ($T_N$ of $NdCrO_3 \sim 220$ K, $T_N$ of $NdFeO_3 \sim 702$ K). Using Eqs. (3), (4) and (5) concurrently, the magnetization at each temperature can be calculated (5). To fit the experimental data using molecular field theory, we employ the most well-known heuristic algorithm—the Marine Predator Algorithm (MPA) [30].

Given that the bonds angle between ions varies with the ratio of $Cr^{3+}/Fe^{3+}$, one should expect the exchange coupling constant varies as well. Let us assume the exchange rate is changed by no more than 10 %, and therefore a precise exchange rates between $0.9J_{ij}$ and $1.1J_{ij}$ can be extracted. Thus, the parameters $J_{NdFe}$, $J_{NdCr}$, and $J_{FeCr}$ are determined using the best fit of the experimental curve. However, although we perform numerous calculations to find appropriate parameters, the fitting result is not reaching the complete picture. Indeed, molecular field theory is an oversimplified model that does not account for Dzyaloshinskii-Moriya (DM) interaction, spin reorientation, or anisotropy. To illustrate the effect of spin reorientation on the magnetism of $NdFe_{1-x}Cr_xO_3$, we consider the magnetization of $NdFe_{1-x}Cr_xO_3$ as the vector superposition of $Fe^{3+}/Cr^{3+}$ and $Nd^{3+}$ ions. Due to the fact that A and B sites have a different easy axis orientation, their interaction can be determined by their spin projection on a particular plane.

In this work, we consider the interacting plane can be ascribed to the Cr/Fe spin, therefore we able to demonstrate how $S_{Nd}$ changes with temperature variation. $S_N$ is defined as the projection of the $Nd^{3+}$ ions' spin on the plane containing the Cr/Fe spin. The exchange constant is fixed across the entire temperature range, which solves the $S_N$ at each temperature. We begin by fitting the magnetization curve of $NdFe_{0.9}Cr_{0.1}O_3$, (Fig. 6a). It turns out that the fitting result is nearly perfect. The $S_N$ is shown in Fig. 6b as a function of temperature. $S_N$ gradually increases as the temperature drops from 400 K to 220 K. The highest $S_N$ is found when the temperature between 100 and 220 K. It indicates that between 100 and 220 K, the B site ion has a binding effect on the A site ion. When the temperature falls below 100 K, $S_N$ begins to decrease with the dropping temperature. Noting that 100 K is the spin reorientation temperature of the system. We convert the $S_N$ change to the angle between the $S_N$ and the Cr/Fe plane in Fig. 6b. The angle increases monotonously as the temperature decreases (a plateau with an angle value of 0 appears between temperatures from 100 K to 220 K). The whole process of spin reorientation can be described in Fig. 8a: when the field of view faces the $Fe^{3+}/Cr^{3+}$ ion spin plane, the magnetic moment of $Nd^{3+}$ rotates from the purple direction (90°) to the gray direction (-90°). This model predicts the phase transition temperature of $NdFe_{0.9}Cr_{0.1}O_3$, to be 695 K in Fig. 6a.

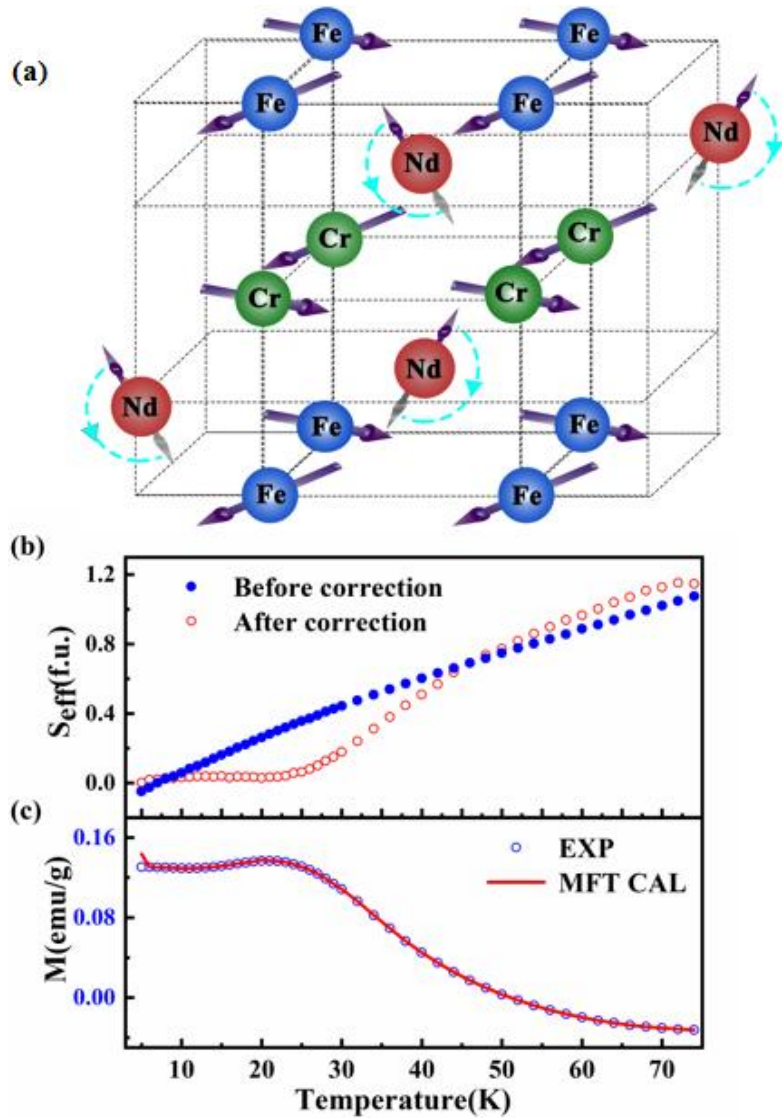

**Fig. 8** (a) Spin reorientation model of NdFe$_{1-x}$Cr$_x$O$_3$ system, (b) the variation of Nd$^{3+}$ spin with temperature before and after correction, and (c) fitting results of NdFe$_{0.1}$Cr$_{0.9}$O$_3$ before and after correction at low temperature.

To analyze further, we substitute the previously obtained spin data into NdFe$_{0.1}$Cr$_{0.9}$O$_3$, and the fitting data obtained are shown in Fig. 6a. It can be seen that the fitted data shows very good results in the high temperature region, and the trend is roughly the same in the low temperature region. Therefore, we consider that the relationship of Nd$^{3+}$ spin with temperature is in agreement to the proposed hypothesis.

The model seen in Fig. 8a can directly describe the spin reorientation process in the NdFe$_{1-x}$Cr$_x$O$_3$ system: the horizontal plane is the spin plane of Fe$^{3+}$/Cr$^{3+}$ ions, and the arrow direction correspond to the magnetic moment of each ion. The direction of the magnetic moment of the Nd$^{3+}$ ion changes with temperature, and its angle with the Fe$^{3+}$/Cr$^{3+}$ spin plane changes from 90° (purple arrow) to -90° (grey arrow). We correct the spin behavior of Nd$^{3+}$ ions at low temperature and the corresponding thermomagnetic curves of NdFe$_{0.1}$Cr$_{0.9}$O$_3$ are shown in Fig. 8(b-c), respectively. It is observed that the curve of spin reorientation represent by the trend after correction is similar to the trend before correction, indicating that the system maintains a similar process of spin reorientation throughout the aforementioned processes. Here we outline the difference is found from the modified curve in which the spin close to 0 at about 25 K, which corresponds to the turning point magnetization at 25 K in the NdFe$_{0.1}$Cr$_{0.9}$O$_3$ thermomagnetic curve. According to previous research, the phase transition temperature of Nd$^{3+}$ ions at the A site in the NdFe$_{1-x}$Cr$_x$O$_3$ system is approximately 1.25 K [31], while the coupled phase transition temperature of Nd$^{3+}$-Fe$^{3+}$/Cr$^{3+}$ is much higher than this temperature. As a result, the magnetization transition around 25 K is realized due to the Nd$^{3+}$-Fe$^{3+}$/Cr$^{3+}$ coupling phase transition process. Therefore, we rationalize the magnetization transition phenomenon is originated from the phase transition driven by the spin reorientation. In other words, the change in the modified effective spin is fairly contributes to the magnetization as well as the phase transition process. The fitting results of exchange constants are shown in Table. 3.

**Table. 3** The fit result of exchange constants in NdFe$_{0.9}$Cr$_{0.1}$O$_3$ and NdFe$_{0.1}$Cr$_{0.9}$O$_3$.

| Sample | $J_{Cr-Fe}$ (K) | $J_{Fe-Fe}$ (K) | $J_{Cr-Cr}$ (K) | $J_{Nd-Fe}$ (K) | $J_{Nd-Cr}$ (K) |
|---|---|---|---|---|---|
| NdFe$_{0.9}$Cr$_{0.1}$O$_3$ | -8.98 | -22 | -15.57 | -2.72 | -1.32 |
| NdFe$_{0.1}$Cr$_{0.9}$O$_3$ | -8.05 | -22 | -14.46 | -4.46 | -3.01 |

## 5. Conclusion

In summary, we have verified the magnetic properties of NdFe$_{1-x}$Cr$_x$O$_3$ with different Cr$^{3+}$/Fe$^{3+}$ ratios in the mechanism of spin reorientation based on a four-sublattice molecular field theory and magnetic measurements. In this study, the projection of Nd$^{3+}$ on the plane can be manipulated by controlling the direction of Cr$^{3+}$/Fe$^{3+}$ magnetic moments via the effective coupling angular momentum. Moreover, the angle between Nd$^{3+}$ and Cr$^{3+}$/Fe$^{3+}$ moments can be tailored under the varying temperature and effective coupling angular momentum variation. As the temperature increases, the effective coupling angular momentum of Nd$^{3+}$ decreases monotonically. The fitting results calculated by the Marine Predator Algorithm are in agreement to the magnetic measurements. The analysis results of this work provide a reference for future researchers to study the magnetic mechanism of other magnetic rare-earth perovskites.

## 6. Conflicts of interest

There are no conflicts to declare.

## 7. Acknowledgment

This work was supported by National Natural Science Foundation of China

(grant number 12105137, 62004143), the Central Government Guided Local Science and Technology Development Special Fund Project (2020ZYYD033), the National Undergraduate Innovation and Entrepreneurship Training Program Support Projects of China, the Natural Science Foundation of Hunan Province, China (grant number S202110555177), the Natural Science Foundation of Hunan Province, China (grant number 2020JJ4517), Research Foundation of Education Bureau of Hunan Province, China (grant number 19A433, 19C1621).

**References**

[1] I. Fita, A. Wisniewski, R. Puzniak, E. E. Zubov, V. Markovich, G Gorodetsky, Phys. Rev. B 98 (2018) 094421.

[2] N. Locatelli, V. Cros, J. Grollier, Nat. Mater. 13 (2014) 11.

[3] R. D. Averitt, A. I. Lobad, C. Kwon, S. A. Trugman, V. K. Thorsmølle, A. J. Taylor, Phys. Rev. Lett. 87 (2001) 017401.

[4] J. Nishitani, K. Kozuki, T. Nagashima, M. Hangyo, Appl. Phys. Lett. 96 (2010) 221906.

[5] A. V. Kimel, A. Kirilyuk, P. A. Usachev, R. V. Pisarev, A. M. Balbashov, Th. Rasing, Nature 435 (2005) 655.

[6] A. V. Kimel, A. Kirilyuk, A. Tsvetkov, R. V. Pisarev, Th. Rasing, Nature 429 (2004) 850.

[7] J. A. de Jong, I. Razdolski, A. M. Kalashnikova, R. V. Pisarev, A. M. Balbashov, A. Kirilyuk, Th. Rasing, A. V. Kimel, Phys. Rev. Lett. 108 (2012) 157601.

[8] P. Jiang, L. Bi, D. H. Kim, G. F. Dionne, C. A. Ross, Appl. Phys. Lett. 98 (2011)

231909.

[9] M. J. Koponen, T. Venalainen, M. Suvanto, K. Kallinen, T-J. J. Kinnunen, M. Harkonen, T. A. Pakkanen, J. Mol. Catal A-Chem. 258 (2006) 246.

[10] D. J. Deka, J. Kim, S. Gunduz, M. Aouine, J-M. M. Millet, A. C. Co, U. S. Ozkan, Appl. Catal. B-Environ. 286 (2021) 119917.

[11] A. Jun, S. Yoo, Y-W. Ju, J. Hyodo, S. Choi, H. Y. Jeong, J. Shin, T. Ishihara, T-H. Lim, G. Kim, J. Mater. Chem. A 3 (2015) 15082.

[12] M. Guo, T. Xia, Q. Li, L. Sun, H. Zhao, J. Eur. Ceram Soc. 41 (2021), 6531.

[13] E. Y. Vedmedenko, H. P. Oepen, A. Ghazali, J. C. S. Lévy, J. Kirschner, Phys. Rev. Lett. 84 (2000) 5884.

[14] A. M. Vibhakar, D. D. Khalyavin, P. Manuel, J. Liu, A. A. Belik, R. D. Johnson, Phys. Rev. Lett. 124 (2020) 127201.

[15] D. C. Johnston, R. J. McQueeney, B. Lake, A. Honecker, M. E. Zhitomirsky, R. Nath, Y. Furukawa, V. P. Antropov, Y. Singh, Phys. Rev. B 84 (2011) 094445.

[16] S.-i. Ohkoshi, Y. Abe, A. Fujishima, K. Hashimoto, Phys. Rev. Lett. 82 (1999) 1285.

[17] J. Shanker, R. V. Kumar, G. N. Rao, D. S. Babu, Mater. Chem. Phys. 251 (2020) 123098.

[18] T. V. Aksenova, Sh. I. Elkalashy, A. S. Urusova, V. A. Cherepanov, Russ. J. Inorg. Chem+ 62 (2017) 1090.

[19] J. Shanker, R. V. Kumar, G. N. Rao, G. N. Babu, Mater. Chem. Phys. 251 (2020) 123098.

[20] T. Shalini, J. Kumar, Bull. Mater. Sci. 44 (2021) 90.

[21] M. Nakhaei, D. S. Khoshnoud, Phys. B Condens. Matter 612 (2021) 412899.

[22] T. Bora, S. Ravi, J. Magn. Magn. Mater. 386 (2015) 85.

[23] J. Shanker, B. V. Prasad, M. B. Suresh, R. V. Kumar, D. S. Babu, Mater. Res. Bull. 94 (2017) 385.

[24] J. J. Mo, R. Y. Xia, Q. H. Zhang, H. W. Chen, L. B. Liu, Y. F. Xia, M. Liu, Phy. Rev. B 105 (2022) 094411.

[25] J. Shanker, M. B. Suresh, G. N. Rao, D. S. Babu, J. Mater. Sci. 54 (2019), 5595.

[26] B. Rajeswaran, D. I. Khomskii, A. K. Zvezdin, C. N. R. Rao, A. Sundaresan, Phys. Rev. B 86 (2012) 214409.

[27] F. A. Fabian, C. C. S. Barbosa, J. G. Santos, R. J. Caraballo-Vivas, F. Garcia, J. G. S. Duque, C. T. Meneses, J. Alloy. Compd. 815 (2020) 152427.

[28] L. Wang, S. Wang, X. Zhang, L. Zhang, R. Yao, G. Rao, J. Alloys Compd. 662 (2016) 268.

[29] S. Lei, L. Liu, C. Wang, C. Wang, D. Guo, S. Zeng, B. Cheng, Y. Xiao, L. Zhou, J. Mater. Chem. A 1 (2013) 11982.

[30] A. Faramarzi, M. Heidarinejad, S. Mirjalili, A. H. Gandomi, Expert. Syst. Appl. 152 (2020) 113377.

[31] M. H. Mohammed, Z. Cheng, S. Cao, J. Horvat, Phys. Chem. Chem. Phys. 23 (2021) 5415.